\documentstyle[12pt]{article}
\title{Fermi co-ordinates and relativistic effects in non-inertial
frames}
\author{Hrvoje Nikoli\'c  \\
Theoretical Physics Division, Rudjer Bo\v{s}kovi\'{c} Institute, \\
P.O.B. 180, HR-10002 Zagreb, Croatia \\
{\normalsize hrvoje@faust.irb.hr} \\
\makebox[1in]{} \\
}
\date{\today}
\begin{document}
\maketitle
\begin{abstract}
Fermi co-ordinates are proper co-ordinates of a
local observer determined by his trajectory in space-time. 
Two observers at different positions belong to different 
Fermi frames even if there is no relative motion between them. 
Use of Fermi co-ordinates leads to several physical 
conclusions related to relativistic effects seen by observers 
in arbitrary motion. In flat space-time, the relativistic 
length seen by an observer depends only on his instantaneous 
velocity, not on his acceleration or rotation. 
In arbitrary space-time, for any observer
the velocity of light is isotropic and 
equal to $c$, provided that it is measured by propagating a light
beam in a small neighbourhood of the observer. The value of a 
covariant field measured at the position of the observer depends 
only on his instantaneous position and velocity, not on his 
acceleration. The notion of radiation is observer independent. 
A ``freely" falling charge in    
curved space-time does not move along a geodesic and therefore 
radiates.    
\end{abstract}

\section{Introduction}

In this paper I review some recent results which have led to  
progress in the understanding of relativistic effects seen by local 
observers arbitrarily moving in flat or curved space-time. 
Some technical details given in the cited papers are omitted here, 
but some conceptual details are explained in a slightly 
different and perhaps more illuminating way.  
I discuss only classical effects, not 
quantum effects, because quantum physics in 
non-inertial frames and curved space-time is not yet 
completely understood.
    
The proper co-ordinates of an observer determined by his trajectory in
space-time are the so-called 
Fermi co-ordinates and they should be used in order to 
describe physical effects seen by him. The main point of 
this paper is the fact that two observers at different positions belong to
different
Fermi frames even if there is no relative motion between them. 
As I discuss later, this fact was not recognized in 
many previous papers, which led to several misinterpretations 
and paradoxes. As I show in the paper, the correct interpretation 
of Fermi co-ordinates leads to several results 
which demonstrate that acceleration and rotation are less important 
than it was previously thought, because it appears that for many 
instantaneous physical effects only the instantaneous position and 
velocity are relevant.

\section{Fermi co-ordinates}

In physics, all dynamical equations of motion are certain 
differential equations that describe certain quantities as 
functions of space-time points. Space-time points 
are parametrized by their co-ordinates. It is convenient to 
write the equations of motion (as well as other related 
equations) in a form which is manifestly covariant with 
respect to general co-ordinate transformations. When one 
solves the equations, one must use some specific co-ordinates. 
The covariance provides that one can use any co-ordinates 
he wants, because later he can easily transform the 
results to any other co-ordinates. Therefore, it is convenient 
to choose co-ordinates that simplify the technicalities of the
physical problem considered.  

The general covariance is often interpreted as a statement 
that ``physics does not depend on the co-ordinates chosen".
However, this is not so. The choice of   
co-ordinates is more than a matter of convenience. 
The main purpose of
theoretical physics is to predict what will be {\em observed} under
given circumstances. 
The main lesson we have learned from Lorentz 
co-ordinates is the fact that what an observer observes
(time intervals, space intervals, components of a tensor, ...)
depends on how the observer moves. Lorentz co-ordinates 
are proper co-ordinates of an observer that moves inertially in 
flat space-time. Fermi co-ordinates are the 
generalization of Lorentz co-ordinates to arbitrary motion 
in arbitrary space-time. If one is interested in how 
a physical system  
looks like to a specific observer, one must transform the results to the
corresponding Fermi co-ordinates. 

Fermi co-ordinates are determined by the (time-like) trajectory of 
the observer, by the rotation of the observer with respect to a local 
inertial observer and by the properties of space-time 
itself. The general geometrical construction of Fermi co-ordinates is 
well established \cite{mtw}. Here I present the most important properties 
of Fermi co-ordinates:
\begin{enumerate}
\item Fermi co-ordinates are chosen such that the position of the 
observer is given by $x^{\mu}=(t,0,0,0)$, where $t$ is the time measured 
by a clock co-moving with the observer.
\item The metric expressed in Fermi co-ordinates has the property
 \begin{equation}
 g_{\mu\nu}(t,0,0,0)=\eta_{\mu\nu} \; .
 \end{equation}
\item The connections $\Gamma^{\alpha}_{\beta\gamma}$ vanish at 
 $(t,0,0,0)$ if and only if the trajectory is a geodesic and 
 there is no rotation.
\end{enumerate}

The general geometrical construction of Fermi co-ordinates is 
not very useful for practical calculations. However, in 
flat space-time, Fermi co-ordinates can be constructed in an alternative way,  
more useful for practical calculations \cite{nels}. 
Here I present a very elegant form of this construction \cite{nikolic1}.
 
Let $S$ be a Lorentz frame 
and let $S'$ be the Fermi frame of the observer whose 3-velocity is
$u^i(t')\equiv \mbox{\bf{u}}(t')$, as seen by an observer in $S$.
In general, $S'$ can also rotate, which can be described by
the rotation matrix $A_{ji}(t')=-A_{j}^{\; i}(t')$.  
The co-ordinate transformation between these two frames is 
given by
\begin{equation}\label{el3}
x^{\mu}=\int_{0}^{t'}f^{\mu}_{\; 0} (t',0;\mbox{\bf{u}}(t')) dt' +
\int_{C}
f^{\mu}_{\; i} (t',\mbox{\bf{x}}';\mbox{\bf{u}}(t')) dx'^{i} \; ,
\end{equation}
where
\begin{equation}\label{partial}
f^{\mu}_{\; \nu}=\left( \frac{\partial f^{\mu}}{\partial x'^{\nu}}
 \right)_{\mbox{\bf{u}}=\mbox{\rm{const}}} \; , 
\end{equation}
and
\begin{equation}\label{el1}
x^{\mu}=f^{\mu}(t',\mbox{\bf{x}}';\mbox{\bf{u}})
\end{equation}
denotes the ordinary Lorentz transformation, i.e. the transformation
between two Lorentz frames specified by the constant relative velocity
$\mbox{\bf{u}}$. Explicitly, 
\begin{eqnarray}\label{fmini}
& f^{0}_{\; 0}=\gamma \; , \;\;\;\;\; f^{0}_{\; j}=-\gamma u_j \; ,
\;\;\;\;\;
  f^{i}_{\; 0}=\gamma u^i \; , & \nonumber \\
& f^{i}_{\; j}=\delta^{i}_{\; j}+
 \displaystyle\frac{1-\gamma}{\mbox{\bf{u}}^2} u^i u_j \; , &
\end{eqnarray}
where $u^j =-u_j$, $\mbox{\bf{u}}^2 =u^i u^i$,
$\gamma=1/\sqrt{1-\mbox{\bf{u}}^2}$ are evaluated at $t'$.
In (\ref{el3}), $C$ is an arbitrary curve with constant
$t'$, starting from $0$
and ending at $-A_{j}^{\; i}(t')x'^{j}$. 
The rotation matrix
satisfies the differential equation
\begin{equation}\label{er4}
 \frac{d A_{ij}}{dt}=-A_{i}^{\; k}\omega_{kj} \; ,
\end{equation}
where $\omega_{ik}=\varepsilon_{ikl}\omega^{l}$, $\varepsilon_{123}=1$  
and $\omega^i(t')$ is the angular velocity as seen by an observer in $S$. 
The Fermi co-ordinates $x'^{\mu}$ 
are constructed such that 
the space origins of $S$ and $S'$ coincide at $t=t'=0$. 
 
The metric tensor in $S'$ is  
\begin{eqnarray}\label{metric}
 & g'_{ij}=-\delta_{ij} \; , \;\;\;\;\;
   g'_{0j}=-(\mbox{\boldmath $\omega$}'\times\mbox{\bf{x}}')_j \; , &
\nonumber \\
 & g'_{00}=c^2 \left(
1+\displaystyle\frac{\mbox{\bf{a}}'\cdot\mbox{\bf{x}}'}{c^2}
    \right)^2 -(\mbox{\boldmath $\omega$}'\times\mbox{\bf{x}}')^2 \; , &
\end{eqnarray}
where
\begin{equation}
 \omega'^i =\gamma (\omega^i -\Omega^i) \; , \;\;\;\;\;
 a'^i =\gamma^2 \left[a^i +\frac{1}{\mbox{\bf{u}}^2}(\gamma
  -1)(\mbox{\bf{u}}\cdot\mbox{\bf{a}})u^i\right] \; ,
\end{equation}
$\Omega^{i}$ is the time-dependent Thomas precession frequency
\begin{equation}
\Omega_{i}=\frac{1}{2\mbox{\bf{u}}^2}(\gamma -1)\varepsilon_{ikj}
 (u^k a^j -u^j a^k) \; ,
\end{equation}
and $a^i=du^i/dt$ is the time-dependent acceleration.
 
In general space-time, instead of a closed formula (\ref{el3}), one can 
find the transformation between two Fermi frames in the form 
\begin{equation}\label{transf}
x^{\mu}=f^{\mu}_{\; \nu} \,
x'^{\nu} + f^{\mu}_{\; \nu\alpha}x'^{\nu}x'^{\alpha}
+ \ldots \; ,
\end{equation}
where it is assumed that the two observers have
the same position at $t=t'=0$. One can show \cite{nikolic2} 
that the quantity 
\begin{equation}\label{fmunu}
f^{\mu}_{\; \nu}=\left( \frac{\partial x^{\mu}}{\partial x'^{\nu}}
 \right)_{x=x'=0} \; 
\end{equation}
is again given by (\ref{fmini}), where $u^i=dx^i/dt$ is the velocity
of the observer in $S'$ as
seen by the observer in $S$, at the instant when the two observers
have the same position.  

From Property 2 we see that the space co-ordinates $x^i$ are a 
generalization of Cartesian co-ordinates. However, this does not 
imply that an observer is not allowed to use polar co-ordinates, 
for example. The most general co-ordinate transformations that 
correspond merely to a redefinition of the co-ordinates of
the same physical observer are the so-called restricted internal transformations 
\cite{nikolic1}, i.e. the transformations of the form 
\begin{equation}\label{veryweak}
t'=f^0 (t) \; , \;\;\;\; x'^i=f^i (x^1,x^2,x^3) \; , 
\end{equation}
where $t,x^i$ are Fermi co-ordinates. The quantities $g_{00}dt^2$ and 
\begin{equation}\label{dl}
dl^2=-g_{ij}dx^{i}dx^{j} \; 
\end{equation} 
do not change under restricted internal transformations. In order 
to describe physical effects as seen by a local observer, one must 
use Fermi co-ordinates modulo restricted internal transformations. 
For simplicity, in the rest of the paper I use Fermi co-ordinates. 

Two observers with different trajectories have different Fermi frames. 
In particular, this implies that {\em even if there is no relative motion 
between two observers, they belong to different frames if they do not 
have the same position}. As we shall see, this fact was not realized 
in many previous papers, which led to many misinterpretations. 
At first sight, this fact contradicts the well-known fact that 
two inertial observers in flat space-time may be regarded as 
belonging to the same 
Lorentz frame if there is no relative motion between them. However, this is 
because their Fermi frames (with the space origins at their positions) 
are related by a translation of the space origin, which is a 
restricted internal transformation. In general, for practical purposes, 
two observers can be regarded  
as belonging to the same Fermi frame if there is no relative motion between
them and the other observer is close 
enough to the first one, in the 
sense that the metric expressed in Fermi co-ordinates 
of the first observer does not depart
significantly from $\eta_{\mu\nu}$ at the position of the second 
one. It is an exclusive
property of Minkowski co-ordinates, among other
Fermi co-ordinates in flat or curved space-time, that the
metric is equal to $\eta_{\mu\nu}$ {\em everywhere}.
This implies that
two observers at different positions but
with zero relative velocity may be regarded as
belonging to the same co-ordinate frame only if they
move inertially in flat space-time.  
  
\section{Relativistic contraction and the rate of clocks}

In this section I study relativistic contraction and the rate of clocks as 
seen by an observer in flat space-time. As seen in the preceding section, 
the explicit co-ordinate transformation that determines 
the relevant Fermi frame is available for flat space-time. 

For motivation, let us first review the problems \cite{nikolic1}
of the standard resolution \cite{gron1,gron2} of the Ehrenfest paradox.  
We study a rotating ring in a rigid non-rotating
circular gutter with radius $r$.
One introduces the co-ordinates
of the rotating frame $S'$
\begin{equation}\label{eq1}
 \varphi'=\varphi -\omega t \; , \;\;\;\; r'=r \; , \;\;\;\; z'=z
  \; , \;\;\;\; t'=t \; ,
\end{equation}
where $\varphi$, $r$, $z$, $t$ are cylindrical co-ordinates of the
inertial frame $S$ and $\omega$ is the angular velocity. The metric in $S'$
is given by
\begin{equation}\label{eq2}
 ds^2=(c^2-\omega^2 r'^2)dt'^2 -2\omega r'^2 \, d\varphi' dt' -dr'^2
 -r'^2 \, d\varphi'^2 -dz'^2 \; .
\end{equation}
It is generally accepted that the space line element should be
calculated by the formula \cite{land}
\begin{equation}\label{eq3}
 dl'^2=\gamma'_{ij}dx'^i dx'^j \; , \;\;\; i,j=1,2,3 \; ,
\end{equation}
where
\begin{equation}\label{eq4}
 \gamma'_{ij}=\frac{g'_{0i}g'_{0j}}{g'_{00}}-g'_{ij} \; .
\end{equation}
This leads to the circumference of the ring 
\begin{equation}\label{eq5}
 L'=\int_{0}^{2\pi} \frac{r' d\varphi'}{\sqrt{1-\omega^2 r'^2/c^2}}
   =\frac{2\pi r'}{\sqrt{1-\omega^2 r'^2/c^2}}   
   \equiv\gamma(r')2\pi r' \; .
\end{equation}
The circumference of the same ring as seen from $S$ is $L=2\pi r=2\pi r'$.
Since the ring is constrained  to have the same radius $r$ as the same 
ring when it does not rotate, $L$ is not changed by the rotation, but
the proper circumference $L'$ is larger than the proper circumference of
the non-rotating ring. This implies that there are tensile stresses in the
rotating ring. The problem is that it is assumed here that (\ref{eq1}) 
defines the proper frame of the whole ring. This
implies that an observer on the ring sees that the 
circumference is $L'=\gamma L$. The circumference of the gutter
seen by him 
cannot be different from the circumference of the ring
seen by him, so the 
observer on the ring sees that the circumference of the
relatively moving gutter is {\em larger} than the proper
circumference of the gutter, whereas we expect that he should
see that it is smaller. 

This problem resolves when one realizes that (\ref{eq1}) does not define 
the proper frame of the ring. Each point on the ring belongs to 
a different Fermi frame. The co-ordinates (\ref{eq1}) are actually 
Fermi co-ordinates (modulo a restricted internal transformation) of an 
observer that rotates in the centre of the ring. 
However, this raises another 
problem. If (\ref{eq3}) is the correct definition
of the space line element, then the observer that rotates in the centre 
should see that the circumference
of the gutter is larger 
than the proper circumference of the gutter by a factor $\gamma(r')$.
However, $\omega r'/c$ can be 
arbitrarily large, so $\gamma(r')$ can be not only 
arbitrarily large, but also even imaginary. On the other hand, we know from
everyday experience that the apparent velocity $\omega r'$ of stars,
due to our rotation,  
can exceed the velocity of light, but we see neither a contraction 
nor an elongation of the stars observed. This implies that the 
definition of the space line element (\ref{eq3}) is not always 
appropriate. In \cite{land}, (\ref{eq3}) 
is derived by defining the space line element 
through a measuring procedure that lasts a finite time, so, in general, this 
formula is not appropriate for a definition of the {\em instantaneous} 
length. 
In flat space-time, as shown in \cite{nikolic1},  
if physics is described by Fermi co-ordinates modulo restricted internal
transformations, a more appropriate definition of the space line element 
is (\ref{dl}).  
 
Using the co-ordinate transformation (\ref{el3}), 
one can study the relativistic contraction in the same way 
as in the conventional approach with Lorentz frames. One assumes that 
two ends of a body are seen to have the same time co-ordinate. From 
(\ref{el3}) and (\ref{fmini}) one can easily see that the co-ordinate 
transformation is linear in $x'^i$. As demonstrated in more detail in 
\cite{nikolic1,nikolic3}, it implies that {\em an arbitrarily accelerated and
rotating
observer sees equal lengths of other differently
moving objects as an inertial observer whose
instantaneous position and velocity are equal to that of the
arbitrarily accelerated and rotating observer.}
  
Using (\ref{el3}), one can also study the rate of clocks as seen by 
various observers. In particular, one can study the twin paradox for 
various motions of the observers. However, it is more interesting to 
study not only the time shift after the two differently moving 
observers eventually meet, but also the continuous changes of the 
time shifts during the motion. As can be seen from (\ref{el3}), 
it is the time dependence (not the space dependence) of the 
co-ordinate transformation that significantly differs from 
the ordinary Lorentz transformations. One cannot invert (\ref{el3}) 
simply by putting $u^i \rightarrow -u^i$. Therefore, inertial 
and non-inertial observers see quite different continuous changes of the 
time shift. 

Here I present the results for uniform circular motion, derived in   
\cite{nikolic1}. Assume that there are three clocks. One is at rest in $S$, 
so it moves inertially. The other two are moving around a circle with the 
radius $R$ and have the velocity $\omega R$ in the 
counter-clockwise direction, as seen by the observer in 
$S$. The relative angular distance between these two non-inertial clocks 
is $\Delta\varphi_0$, as seen in $S$. The inertial observer will see that 
the two non-inertial clocks lapse equally fast, so I choose that he sees 
that they show the same time. He will see the clock rate $t=\gamma t'$, 
where $\gamma=1/\sqrt{1-\omega^2 R^2/c^2}$. The observer co-moving with 
one of the non-inertial 
clocks will not see that the other non-inertial clock shows the 
same time as his clock. 
He will see the constant time shift given by the equation
\begin{equation}\label{eqgron1}
\gamma\omega (t''-t')=\beta^2\sin (\gamma\omega
(t''-t')+\Delta\varphi_0) \; ,
\end{equation}
where $\beta^2 \equiv \omega^2 R^2 /c^2$ and $t''$ is the time of the other 
non-inertial clock.
Finally, let us see how the inertial clock looks like from the point 
of view of the observer co-moving with one of the non-inertial clocks. 
Let the position of the inertial clock be given by its co-ordinates 
$(x,y)$. We choose the origin of $S$ such that, at $t=t'=0$, the 
space origins of $S$ and $S'$ coincide  
and the velocity of the non-inertial 
observer is in the $y$-direction, as seen in $S$. The rate of 
clocks as seen by the non-inertial observer is given by
\begin{equation}\label{txy}
t=\gamma t' + 
 \frac{\omega R}{c^2}[y \cos \gamma \omega t'
 -(x+R) \sin \gamma \omega t'] \; .
\end{equation}
The oscillatory functions in (\ref{txy}) vanish when
they are averaged over time. 
This means that the observer in $S'$ agrees
with the observer in $S$ that the clock in $S'$ is slower, but only in
a time-averaged sense. For example, when these two clocks 
are very close to each other, 
then, by expanding (\ref{txy}) for small $t'$, one finds $t=t'/\gamma$, 
which is the result that one would obtain if the velocity of the
non-inertial  
observer were constant.  
If the clock in $S$ is in the centre, which
corresponds
to $x=-R$, $y=0$, then (\ref{txy}) gives $t=\gamma t'$, so in this case
there is no oscillatory behaviour.     

\section{The invariance of radiation and of the velocity of light}

In this section I study some physical effects for which an explicit 
transformation, such as (\ref{el3}), is not necessary. This allows 
to drew conclusions which refer to general (not only  
flat) space-time. 

If (\ref{eq1}) is interpreted as a proper frame of all observers on a 
rotating platform, then one can conclude that the observer on the 
rotating platform will see that the velocity of light depends on 
whether light is propagating in the clockwise or in the counter-clockwise 
direction (see, for example, \cite{kla}). This is related to the 
fact that the metric (\ref{eq2}) is not time orthogonal.  
However, now we know that each observer
belongs to a different Fermi frame, and from Property 2 we see
that
in the {\em vicinity} of any observer the metric is equal to the
Minkowski metric $\eta_{\mu\nu}$. This implies that {\em
for any local observer the velocity of light is isotropic and is
equal to $c$, provided that it is measured by propagating a light
beam in a {\bf small} neighbourhood of the observer.} 
In particular, this leads to a 
slightly different interpretation of the Sagnac effect \cite{nikolic1}. 

The result that (\ref{fmunu}) is given by (\ref{fmini}) has the 
following very general implication. Let $\Phi_{\alpha_1 \ldots \alpha_n}(x)$
be
an arbitrary local tensor quantity. Let the two observers
measure this quantity at their common instantaneous position.
The results of measurements will be related as
\begin{equation}\label{Phi}
\Phi'_{\mu_1 \ldots \mu_n}=f^{\alpha_1}_{\; \mu_1}
 \cdots f^{\alpha_n}_{\; \mu_n} \, \Phi_{\alpha_1 \ldots \alpha_n} \; .
\end{equation}
The two measurements will be different if there is an instantaneous
relative velocity between the two observers. However, the 
instantaneous relative acceleration, as well as higher-order 
derivatives, are irrelevant to this transformation law.

A local observer can measure only
the values of fields at the point of his own position.
It is completely
unphysical to talk about the value of a field at some point
from the point of view of the observer sitting at some other
point. Since all interactions are local, a measuring 
apparatus can respond only to the values of fields at the position
of the apparatus. This implies that {\em covariant fields seen by 
an observer 
depend only on his instantaneous velocity, not on his acceleration}.

This is in contrast with what was concluded in several previous 
papers (see references in \cite{nikolic2}). 
In particular, it was concluded that an inertially moving charge
radiated
from the point of view of an accelerated observer and that the
accelerated charge did not radiate from the point of view of an
co-accelerating observer. These conclusions were an artifact of the 
misinterpretation of the Fermi co-ordinates of the accelerated observer, 
in the sense that they were applied to calculate how the fields 
transformed at {\em all} points covered by these co-ordinates, not only 
at the point of the position of the observer.  

The use of Fermi co-ordinates also illuminates the origin of radiation 
of an accelerated charge. The radiation is not a kinematical effect
resulting from the co-ordinate transformation between
the frames of the radiating charge and the observer, but a
dynamical effect, in the sense that even for the observer
co-moving with the charge, the fields depend on acceleration.
This is not explicitly seen in the conventional approach  
in which the Maxwell equations are solved in
Minkowski co-ordinates. To see
this explicitly, we write the covariant Maxwell equation 
\begin{equation}\label{max1}
D_{\mu}F^{\mu\nu}=j^{\nu} 
\end{equation}
in a more explicit form 
\begin{equation}\label{max2} 
\partial_{\mu}F^{\mu\nu}+\Gamma^{\mu}_{\mu\lambda}F^{\lambda\nu}
 +\Gamma^{\nu}_{\mu\lambda}F^{\mu\lambda}=j^{\nu} \; .
\end{equation}  
We assume that the current $j^{\nu}$ corresponds to a point-like
charge. Let us study how the field $F^{\mu\nu}$ looks like to
an observer co-moving with the charge in his small neighbourhood.
Since he uses the corresponding Fermi co-ordinates,
the connections $\Gamma^{\alpha}_{\beta\gamma}$
vanish in his
small neighbourhood if and only if his trajectory is a
geodesic (I assume that the charge does not rotate). Therefore, 
if the charge does not accelerate, in the small neighbourhood
of the charge the solution of (\ref{max2}) looks just like  
the well-known Coulomb solution ${\bf E}\!\propto\! r^{-2}$,
${\bf B}\! =\! 0$. 
On the other hand, if the charge accelerates, then, even in the
small neighbourhood, Eqs. (\ref{max2}) no longer look     
like the Maxwell equations in Minkowski space-time.
This gives rise to a more complicated solution, which includes the
terms proportional to $r^{-1}$. The essential feature of radiating 
fields is the fact that they fall off with distance 
much slower than other fields, so their effect is 
much stronger at large distances. If the field is proportional
to $r^{-1}$ as seen by one observer, it is also so as seen by any other
observer at the same position. This means that the notion of radiation 
does not depend on the observer. 

As is well known, when a charge accelerates, a self-force acts on it. 
A self-force also appears when a charge moves geodesically in curved 
space-time \cite{hobbs,qwald}. 
Fermi co-ordinates provide a simple intuitive picture explaining why
a self-force  
appears when the charge accelerates or when space-time around the charge
is curved. The fields produced by the charge always act on it. However,
when the charge moves inertially through flat space-time, then   
the metric related to the corresponding Fermi co-ordinates is
isotropic,
and so are the fields. This implies that self-forces
in different directions cancel exactly, so the resultant force
is zero. When space-time is curved (such that it is not isotropic)
or the charge accelerates, then 
the metric related to the Fermi co-ordinates is no longer 
isotropic. Consequently, the fields are also not isotropic, 
which implies 
that the resultant force need not to be zero. This implies that  
{\em a ``freely" falling charge in    
curved space-time does not move along a geodesic and therefore 
radiates}. 

\section*{Acknowledgment}

This work was supported by the Ministry of Science and Technology of the
Republic of Croatia under the contract No. 00980102.

\end{document}